\documentstyle[eqsecnum,aps]{revtex}

\begin{document}
\draft
\preprint{HEP/123-qed}
\title{Path Integral Approach to Resistivity Anomalies in Anharmonic Systems}
\author{Marco Zoli}
\address{Istituto Nazionale di Fisica della Materia - 
Dipartimento di Matematica e Fisica, \\  Universit\'a di Camerino, 
62032 Camerino, Italy. e-mail: zoli@campus.unicam.it
}

\date{\today}
\maketitle
\begin{abstract}
Different classes of physical systems with sizeable
electron-phonon coupling and lattice distortions
present anomalous resistivity behaviors versus 
temperature. We study a molecular lattice Hamiltonian
in which polaronic charge carriers interact with
non linear potentials
provided by local atomic fluctuations between
two equilibrium sites.
A path integral model is developed to select the class of
atomic oscillations which mainly contributes to the
partition function and the electrical resistivity is computed 
in a number of representative cases.
We argue that the common origin of the observed 
resistivity
anomalies lies in the time retarded nature of 
the 
polaronic interactions in the local
structural instabilities.
\end{abstract}

\narrowtext
\section*{I.Introduction}

While a sizeable electron-phonon coupling is
prerequisite for polaron formation, it is not generally
true that a system with strong electron-phonon interactions
($g > 1$) has to display polaronic features.
In a simple metal with wide bare bands and long range
forces like lead, for
instance, intersite off-diagonal hopping terms prevent
trapping of the charge carriers and onset of a polaronic state. 
In  strong coupling systems with a narrow electronic band 
$D$ the antiadiabatic
inequality $D/\hbar \bar \omega < 1$ ($\bar \omega$ being the 
characteristic
phonon frequency) is likely fulfilled thus ensuring the stability
of the unit comprising electron and phonon cloud that is, 
the small
polaron \cite{lang}. Also in adiabatic regimes small polarons 
can form although
a real space broadening of the quasiparticle has to be expected, 
at least
in three dimensional systems. As a consequence, the adiabatic 
polaron mass renormalization
is less strong than that occuring for the antiadiabatic polaron
once the strength of the electron-phonon coupling has been fixed \cite{eminholst}.
Theoretical investigation on these issues has become
intense after that a polaronic mechanism has been proposed 
\cite{alemott} as a viable
possibility to explain high $T_c$ superconductivity with its
peculiar transport properties. Evidence has been recently 
provided \cite{zhao} for the polaronic nature of carriers also 
in perovskite manganites
with colossal magnetoresistance (CMR) 
and a strong coupling of 
carriers to Jahn-Teller
lattice distortions has been suggested \cite{millis} to 
explain the resistivity peak located at the Curie temperature \cite{schiffer}.
Although polarons are invoked to account for different
resistivity behaviors in systems with some lattice 
distortions \cite{bianconi}, the common origin, 
if any does exist, of
such behaviors remains unexplained and a unifying theory
is lacking. In general, a strong electron-phonon coupling
implies violation of the Migdal theorem \cite{yuand} 
and polaron
collapse of the electron band with the appearance of
time retarded interactions in the system. The path
integral method \cite{feynman,devreese} has proved successful 
in dealing with
this problem since a retarded potential naturally
emerges in the exact integral action. Both the 
Fr\"ohlich \cite{frohlich,ganbold} and the Holstein \cite{holstein,deraedt,kornilo}
polaron properties have been studied by using
path integrals techniques.

In this paper we propose a path integral approach to
the problem of a polaron scattered by a local
lattice instability arising from an anharmonic phonon
mode. Small polarons are assumed to exist by virtue
of the strong {\it overall} electron-phonon coupling
and independently of the {\it single} anharmonic mode.
As an example, the latter could be associated with the
motion of apical oxygen atoms in $YBa_2Cu_3O_7$ superconductor 
where a c-axis polarized high frequency
phonon couples to in-plane charge carriers \cite{timusk}.
On the other hand, the highly correlated motion of in-plane 
holes and c-axis lattice vibrations could in {\it itself} 
give origin to a 
small polaron \cite{mustre}.
We  model the local instability by a double well
potential in its two state configuration, a Two Level
System (TLS), and the one dimensional atomic path between
the two equilibrium minima is taken as time dependent
\cite{takeno}.
Retardation effects are thus naturally introduced 
in the full partition function and the path integral 
method permits to derive the effective time (temperature)
dependent coupling strengths which control the resistivity.

\section*{II. The Path Integral Model}

Our starting point is contained in the following
Hamiltonian

\begin{eqnarray}
H_{0}(\tau)=& &\, \bar \epsilon(g) \tilde c^{\dag}(\tau) 
\tilde c(\tau) + \sum_{\bf q}{\omega_{\bf q}}a^{\dag}_{\bf q}
(\tau)a_{\bf q}(\tau) + H_{TLS}(\tau) \,
\nonumber \\
& &\Bigl( H_{TLS}(\tau) \Bigr)=\, 
\left(\matrix{0 & \lambda Q(\tau) \cr 
\lambda Q(\tau) & 0 \cr} \right)\,
\nonumber \\  
& &H_{int}(\tau)=\,- 2\lambda Q(\tau) \tilde c^{\dag}(\tau) 
\tilde c(\tau) \,
\nonumber \\
& &Q(\tau)=\, -Q_o + {{2 Q_o}\over {\tau_o}}
(\tau - t_i)
\label{1}.
\end{eqnarray}

$\tau$ is the time which scales as an inverse temperature
according to the Matsubara Green's function formalism.
$H_0(\tau)$ is the free Hamiltonian made of: a) a polaron
created (distroyed) by $\tilde c^{\dag}(\tau) ( \tilde c(\tau))$ 
in an energy band $\bar \epsilon(g)$ whose width decreases 
exponentially by increasing the strength of the overall
electron-phonon coupling constant $g$, 
$\bar \epsilon(g)=\,Dexp(-g^2)$; b) a diatomic
molecular lattice whose phonon frequencies
$\omega_{\bf q}$ are obtained analytically through a
force constant approach; c) a Two Level System  in its 
symmetric ground state configuration due
to an anharmonic phonon mode.
$Q(\tau)$ is the one dimensional {\it space-time}
hopping path followed by the atom which moves between
two equlibrium positions located at $\pm Q_o$.
$\tau_o$ is the bare hopping time between the two minima 
of the TLS and
$t_i$ is the instant at which the $ith$-hop takes place.
One atomic path is characterized by the number $2n$ of hops, 
by the set of $t_i$ $(0 < i \le 2n)$ and by $\tau_o$.
We have implicitly assumed that the class of 
$\tau$-linear paths yields the main contribution to
the full partition function of the interacting system
\cite{hamann}.
The closure condition on the path is given by:
$(2n - 1)\tau_s + 2n\tau_0 =\, \beta$, where  $\beta$ is the inverse temperature and $\tau_s$ is the time one 
atom is sitting in a well. 
The interaction is described by $H_{int}(\tau)$ with
$\lambda$ being the coupling strength between TLS and
polaron, 
$\lambda Q(\tau)$ is the renormalized (versus time) tunneling 
energy which allows one to introduce the $\tau$ dependence in 
the interacting Hamiltonian \cite{zoli}.

Following a method previously developed 
\cite{hamann} in the study of the Kondo problem, 
we multiply $\lambda Q(\tau)$ by a fictitious 
coupling
constant $s$ $(0 \le s \le 1)$ and, by differentiating with 
respect to $s$,
one derives the  one path
contribution to the partition function of the system

\begin{equation}
ln \Biggl({{Z(n,t_i)} \over {Z_0}} \Biggr)=\, -2 \lambda 
\int_0^1 ds \int_0^{\beta} d\tau Q(\tau) 
 \lim_{\tau^{'} \to \tau^+} G(\tau,\tau^{'})_s  
\label{2}
\end{equation}

where, $Z_0$ is the partition function related to $H_{0}$
and $G(\tau, \tau^{'})_s$ is the full propagator for polarons
satisfying a Dyson's equation:

\begin{equation}
G(\tau, \tau^{'})_s=\,
   G^{0}(\tau, \tau^{'}) + s \int_0^{\beta} dy
   G^{0}(\tau, y) \lambda Q(y)
   G(y, \tau^{'}) 
\label{3}
\end{equation}

The polaronic free propagator $G^{0}$ can be derived exactly
in the model displayed by eqs.(1). 
We get the full partition function of the system  
by integrating 
over the times $t_i$ and summing over all possible even 
number of hops:

\begin{eqnarray}
& &Z_T=\, Z_0\sum_{n=0}^{\infty}
\int_0^{\beta}{{dt_{2n}}\over {\tau_0}} \cdot 
\cdot
\int_0^{t_2-\tau_0}{{dt_{1}}\over {\tau_0}}
exp\Bigl[-\beta E(n,t_i, \tau_0) \Bigr]  \,
\nonumber \\
& &\beta E(n,t_i, \tau_0)=\,
\L -
\bigl(K^{A} + K^{R}\bigr) \sum_{i>j}^{2n}
\biggl({{t_i - t_j}\over {\tau_0}}\biggr)^2 
\label{4}
\end{eqnarray}

with $E(n,t_i, \tau_0)$ being the one path atomic energy. 
$\L$, which is a function of the input parameters,
is not 
essential here while the second addendum in eq.(4) is 
{\it not local in time} as a result of the retarded polaronic
interactions between successive atomic hops in the
double well potential. $K^{A}$ and $K^{R}$ are the one path
coupling strengths containing the physics of the interacting 
system. $K^{A}$ (negative) describes the 
polaron-polaron attraction mediated by
the local instability and $K^{R}$ (positive)
is related to to the repulsive scattering of the polaron
by the TLS. Computation of $E(n,t_i, \tau_0)$ and its
derivative with respect to $\tau_0$ shows that the largest
contribution to the partition function is given by the
atomic path with $\tau_s=\,0$. The atom moving back and
forth in the double well minimizes its energy if it takes
the path with
the highest $\tau_0$ value allowed by the boundary
condition, that is with 
$(\tau_0)_{max}=\, (2nK_BT)^{-1}$.
This result, which is general, provides a 
criterion
to determine the set of dominant paths for the atom at any 
temperature. Then, the effective interaction strengths  
$<K^{A}>$ and  $<K^{R}>$
can be obtained as a function of $T$ by summing 
over $n$ the dominant
paths contributions:

\begin{eqnarray}
& &<K^{A}>=\,- {\bigl(\lambda Q_0\bigr)^2 B^2 
exp\bigl({2 \sum_{\bf q}A_{\bf q}}\bigr)
\sum_{\bf q}A_{\bf q} \omega_{\bf q}^2 } 
\tilde f  \,
\nonumber \\
& & <K^{R}>=\,- { \beta \bigl(\lambda Q_0\bigr)^3 B^3 
exp\bigl({3 \sum_{\bf q}A_{\bf q}}\bigr) 
\sum_{\bf q}A_{\bf q} \omega_{\bf q}^2 } 
\tilde f
\label{5}
\end{eqnarray}

with: $B=\,(n_F(\bar \epsilon) - 1) 
exp(- g \sum_{\bf q}ctgh(\beta \omega_{\bf q}/2 ))$ and
$A_{\bf q}=\,2g \sqrt{N_{\bf q}(N_{\bf q} + 1)}$.

$N_{\bf q}$ is the phonon occupation factor and
$n_F(\bar \epsilon(g))$ is the Fermi distribution for
polarons.
$\tilde f =\,\sum_{n=1}^N (\tau_0)_{max}^4$ and
$N$ is the cutoff on the number of hops in a path. 
The particular form of
$(\tau_0)_{max}$ suggests that many hops paths
are the relevant excitations 
at low temperatures whereas paths with a low number
of hops provide the largest contribution to the 
partition 
function at high temperatures. Since the effective
couplings which determine the resistivity depend 
on $(\tau_0)_{max}^4$, hence
on $N^{-3}$ (through $\tilde f$), a
relatively small cutoff ($N \simeq 4$) ensures numerical 
convergence of eqs.(5) in the whole temperature range.
On the other hand the non retarded term $\L$ in eq.(4) 
has a slower
$1/N$ behavior,
therefore a larger cutoff should be taken
at low temperatures where computation of
equilibrium properties such as specific heat is strongly 
influenced by many hops atomic paths between the minima
of the double well excitations.

Our lattice Hamiltonian is made of diatomic sites
whose intramolecular vibrations can favor trapping
of the charge carriers  \cite{holstein}. 
The {\it intra}molecular 
frequency $\omega_0$ largely influences the size of
the lattice distortion associated with  polaron formation
\cite{alenew} while the dispersive features of the phonon
spectrum controlled by the {\it inter}molecular couplings
are essential to compute the polaron properties
both in the ground state and at finite temperatures.
The range of the intermolecular forces is extended to the
second neighbors shell since these couplings 
remove the phonon modes
degeneracy (with respect to dimensionality) at the
corners of the Brillouin zone thus permitting to estimate 
with accuracy the relevant contributions of high symmetry 
points to the momentum space summations.

In a simple cubic lattice model with first and second
neighbors molecular sites interacting via a force
constants pair potential, the $\omega_{\bf q}$ are: 

\begin{eqnarray}
& &\omega^2_{\bf q}= 
{{\alpha + 3 \gamma + 4 \delta} \over M} + 
{1 \over M} \sqrt { \alpha^2 +  F_{x,y,z}} \,
\nonumber \\
& &F_{x,y,z}= \gamma ^2 (3 + 2(c_xc_y + s_xs_y + 
c_xc_z + s_xs_z + c_yc_z   \,
\nonumber \\
& & + s_ys_z) ) + 2 \alpha \gamma (c_x + c_y + c_z) +
4 \alpha \delta (c_x c_y + c_y c_z ) \,
\nonumber \\
& & + 2 \gamma \delta ( 3c_y + 2c_z +  c_x
   + 2cos(q_x - q_y - q_z) \,
\nonumber \\
& & + 2c_x (c_yc_z + s_xs_z) + c_xc_{2y} + s_xs_{2y} +
   c_yc_{2z} + s_ys_{2z}) \,
\nonumber \\
& & + 2 \delta^2 (2 + 2c_x c_z +   
   2c_z(c_xc_{2y} + s_xs_{2y}) + c_{2y} + c_{2z} ) \,
\label{6}
\end{eqnarray}

where, $c_x=cosq_x$, $s_x=sinq_x$, $c_{2y}=cos2q_y$ etc.
The intramolecular force constant $\alpha$ is related 
to $\omega_0$ by  $\omega_0^2=\,2\alpha/M$
with $M$ being the reduced molecular mass.
$\gamma$ and $\delta$ are the intermolecular first  
and second neighbors force constants, respectively. 
Let's define $\omega_1^2=\,\gamma/M$ and $\omega_2^2=\,\delta/M$.
Then the characteristic frequency $\bar \omega$, 
which we choose as the zone 
center frequency, is: 
$\bar \omega^2=\, \omega_0^2 + z\omega_1^2 
 + z_{nnn}\omega_2^2$, $z$ is the coordination number and 
$z_{nnn}$ is the next nearest neighbors number.

We turn now to compute the electrical resistivity 
due to the polaronic charge carrier scattering by
the impurity potential with internal degree of freedom
provided by the TLS.
Assuming s wave scattering, one gets \cite{mahan,yuand}

\begin{eqnarray}
& &\rho=\, \rho_0 sin^2 \eta \,
\nonumber \\
& &\rho_0=\, {{3 n_s} \over {\pi e^2 v_F^2 
\bigl( N_0/V \bigr)^2 \hbar}}\,
\label{7}
\end{eqnarray}

where, $n_s$ is the density of TLS which act as scatterers,
$v_F$ is the Fermi velocity, $V$ is the cell volume, 
$N_0$ is the electron density of states,
$e$ is the 
charge and $\hbar$ is the Planck constant. The 
phase shift $\eta$ to the electronic wave function at the Fermi surface
is related to the effective interaction strengths 
$<K^{A}>$ and  $<K^{R}>$.

The input parameters of the model are six that is,
the three molecular force constants, $g$, $D$ and the bare
energy $\lambda Q_0$.
$Q_0$ can be chosen as $\simeq 0.05 \AA$ consistently with
reported values in the literature on TLS's which are
known to exist in glassy systems \cite{ander,fessa}, 
amorphous metals \cite{coch},
A15 compounds \cite{yuand} and likely in some cuprate 
superconductors \cite{mustre,saiko}. In these systems the origin
of the TLS's is not magnetic. The bare electronic band
$D$ is fixed at 0.1eV.
In Fig.1, we take a rather large $\bar \omega$ setting the system 
in moderately adiabatic conditions and an intermediate ($g=\,1$) 
to very strong ($g > 3$) electron-phonon
coupling regime. The broad resistivity maximum 
developing at low temperatures in the intermediate regime
clearly signals that the TLS's are at work here \cite{tsuei} 
while, by increasing $T$, the hopping time
shortens and the incoming polarons cannot distinguish
any more the TLS internal degree of freedom, then
diagonal scattering prevails and a {\it quasi linear} 
resistivity behavior emerges at $T > 200K$. At larger $g$ the polaron
becomes progressively smaller in real space and 
the occurence of the self trapping event (at $g \simeq 3$)
is marked by an abrupt increase in the effective mass, in any
dimensionality
\cite{io,romero}. As a consequence, the resistivity becomes
non-metallic and its absolute values are strongly enhanced.
Let's fix $g=\,1$ which ensures polaron mobility and tune 
(see Fig.2) the TLS-polaron coupling $\lambda$. We see
that a resistivity peak located at $T \simeq 150K$ arises
at $\lambda \ge 700meV \AA^{-1}$ with height and 
width of the peak being strongly dependent on $\lambda$
hence, on the TLS energy. The low $T$ resistivity
still displays the maximum at the unitary limit while
the high $T$ ($T > 300K$) behavior can be metallic like (
$\lambda < 800meV \AA^{-1}$) or semiconducting like 
($\lambda > 800meV \AA^{-1}$). 
$\lambda \simeq 700 - 800meV \AA^{-1}$ corresponds to a
TLS energy of $\simeq 35 - 40meV$ which is comparable
to the value of the bare polaron energy
band $\bar \epsilon(g=1) \simeq 37meV$. 
In this picture, the resistivity peak has a structural
origin and it can be ascribed to resonant
TLS-polaron scattering with effective attractive and
repulsive interaction strengths 
becoming of the same
order of magnitude at $T \simeq 150K$. 
While the peak does not shift substantially by varying the strength
of the intermolecular forces the height of the peak turns out to be
rather sensitive to those strength hence, to the size of the quasiparticle.
As an example, by doubling $\omega_1$ with respect to the value in Fig.2 
(at fixed $\omega_0$ and $\omega_2$), the normalized 
resistivity peak ($\lambda=\,1000meV \AA^{-1}$ curve) 
drops to 0.65 while by taking $\omega_1=\,60meV$ the
peak disappears. The effect of the second neighbors force
constant is relatively less strong:
by doubling $\omega_2$ (at fixed $\omega_0$ and $\omega_1$) 
with respect to fig.2, the resistivity
maximum becomes 0.76 while $\omega_2=30meV$ yields a
peak value of 0.66.

At larger $\lambda$'s the
resonance peak is higher and broader since an 
increasing number of incoming
polarons can be off diagonally scattered by the TLS.
However, the appearance of this many body effect 
mediated by
the local potential does not change the position 
of the peak either, which
instead can be shifted towards lower temperatures
by reducing substantially $\omega_0$ 
hence the $\bar \omega$ value. 
In Fig.3, we set $\omega_0=\,50meV$. As previously discussed
\cite{io} the intermolecular coupling $\omega_1$ should be
$\simeq \omega_0/2$ to ensure a correct application of the
molecular lattice model with first neighbors interacting 
sites. Here we are extending the range of the 
forces  to the second neighbors shell accordingly, $\omega_1$
is slightly reduced to the value $20meV$ while $\omega_2$
is switched on and fixed at $10meV$.
This choice of parameters allows us: 
i) to treat correctly the ground state polaron
properties versus dimensionality \cite{io}, 
ii) to set the characteristic phonon frequency of the
present three dimensional lattice model at
$\bar \omega \simeq 75meV$ which is the energy of c-axis 
polarized phonons due to apical oxygen vibrations 
coupled to the holes in the Cu-O planes of $YBa_2Cu_3O_{7-\delta}$ \cite{timusk}. Although our simple cubic lattice does not account
for the details of the structural effects of $YBCO$ we are 
in the
appropriate range of parameters to capture the main features of the
lattice polarons scattered by local instabilities in those compounds.

Anharmonic features of the oxygen modes have been recognized to be 
larger in underdoped samples \cite{palles} and doping dependent 
polaron
formation \cite{haskel} has been correlated to
distortions of the oxygen environment. 
Our model predicts (Fig.3) a large resisitivity peak whose height
is strongly increased with respect to Fig.2 (compare the 
$\lambda=\,700meV \AA^{-1}$ cases)
due to the fact that the polaron 
effective mass is heavier when lower energy phonons build
up the quasiparticle.  
The non metallic behavior at $T$ larger than $\simeq 90K$
reminds of the anomalous c-axis resistivity observed in underdoped
high $T_c$ superconductors \cite{ito}.
We believe that the semiconducting like $\rho_c$
in underdoped high $T_c$ superconductors can be ascribed to 
anharmonic
potentials due to oxygen displacements strongly coupled via 
$\lambda$
(as in Fig.3) to polaronic 
carriers. To attempt a comparison with experiments 
we need to fix the residual resistivity through the
ground state parameters given in eq.(7). By extrapolating to 
$T=\,0$ the normal state data on $YBa_2Cu_3O_{7-\delta}$ one 
derives $\rho(T=\,0) \simeq 3m\Omega cm$, hence the experimental 
peak value $\rho \simeq 20m\Omega cm$ observed in the 80K compound $YBa_2Cu_3O_{6.87}$ \cite{ito} can be reproduced in my model by
$\lambda \simeq 1300meV \AA^{-1}$ which corresponds to a local 
mode energy of $\simeq 65meV$, in fair agreement with the measured 
energies of phonons strongly coupled to the charge carriers.

In the range $80K < T < 200K$ the $\lambda \simeq 1300meV \AA^{-1}$ curve
fits rather well the data of ref. \cite{ito}. At $T=\,200K$, 
the experimental $\rho$ is $\simeq 10m\Omega cm$ 
and the calculated 
value is $\simeq 8m\Omega cm$. In the range 
$200K < T < 300K$, the experimental $\rho$ is essentially flat
while my results still exhibit a $d\rho/dT < 0$ behavior.
In general, the 
resistivities here obtained have an exponential temperature dependence
(at $T$ larger than the resonance) consistent with a polaronic
hopping motion whereas the experimental data show a variety
of sample dependent non metallic behaviors for the out of
plane resistivity \cite{liu} in underdoped compounds.

A resistivity peak at $T \simeq 250K$ is observed in
CMR materials
\cite{schiffer} and a splitting of the Mn
states due to a local lattice distortion has been proposed
as a possible mechanism. In our description, 
only extended polarons
dragging a cloud of high frequency optical phonons (see Fig.4)
can produce a  peak at $T \simeq 200- 250K$
which is however broader than the experimental one.  
High phonon frequencies
imply reduced mass renormalization and, consistently, weak overall
electron-phonon coupling values. Extended polarons move therefore 
in wide energy bands and the resonance effect can take place only
if the TLS splitting energy is large enough. 
Infact, by taking $D=0.5eV$ and $\lambda \simeq 5000meV \AA^{-1}$ 
(Fig.4), we get the peak at $g \le 1$ whereas, 
in a small polaron regime ($g > 1$)
the resonance is lost and the peak is suppressed.
Although more specific models accounting for magnetic  
field effects are required for the CMR systems, 
we argue that
the strong coupling of polarons to structural
distortions is likely relevant to their transport properties.

\section*{III. Conclusion}

The path integral
method provides a powerful tool to describe
{\it non local (in time)}
scattering of the charge carriers by 
{\it local (in space)} potentials due to structural distortions. Our charge carriers are 
lattice polarons which exist in the diatomic molecular lattice by virtue of a
sizeable overall electron-phonon coupling. The structural
instability is due to an anharmonic atomic motion  
between two equilibrium positions which typically could
be $\simeq 0.1\AA$ apart. I have derived the full partition
function of the interacting system and obtained the
temperature dependent effective couplings by summing over
the class of energetically favored atomic hopping paths.
We suggest
that the retarded nature of the
interactions is the {\it key} ingredient
to explain  some
anomalous transport properties observed in real materials.
A resistivity peak is obtained as a consequence of 
resonant scattering of polarons strongly coupled to 
local double well potentials.
\begin{figure}
\vspace*{6truecm}
\caption{Electrical Resistivity normalized to the residual
($T=\,0$) resistivity for five values of electron-phonon coupling
$g$. The bare TLS energy $\lambda Q_0$ is 6.5meV. 
The force constants which control the phonon
spectrum are: $\omega_0=\,100meV, \omega_1=\,20meV,
\omega_2=\,10meV$}
\end{figure}

\begin{figure}
\vspace*{6truecm}
\caption{Electrical Resistivity normalized to the residual
resistivity for five values of the polaron-TLS coupling
$\lambda$. $g=\,1$. The intra- and intermolecular force 
constants  are as in Fig.1}
\end{figure}

\begin{figure}
\vspace*{6truecm}
\caption{Electrical Resistivity normalized to the residual
resisitivity  for nine values of the polaron-TLS coupling
$\lambda$. $g=\,1$.  $\omega_0=\,50meV, \omega_1=\,20meV,
\omega_2=\,10meV$}. 
The experimental data (X) are taken from ref.\cite{ito}.
\end{figure}

\begin{figure}
\vspace*{6truecm}
\caption{Electrical Resistivity normalized to the residual
resistivity for four overall couplings $g$.
$\lambda=\,5000meV \AA^{-1}$.   
$\omega_0=\,100meV, \omega_1=\,80meV,
\omega_2=\,20meV$}
\end{figure}

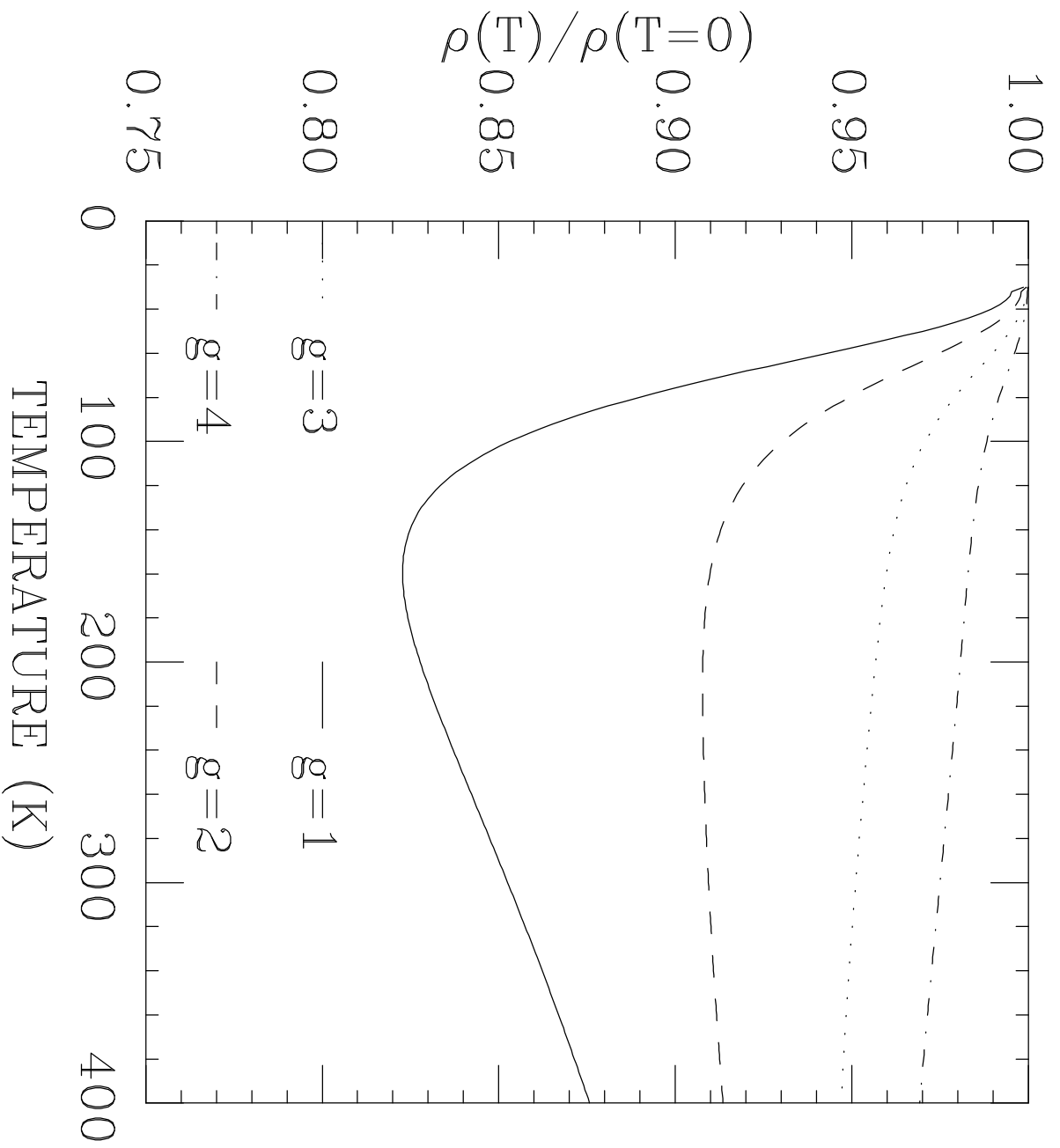
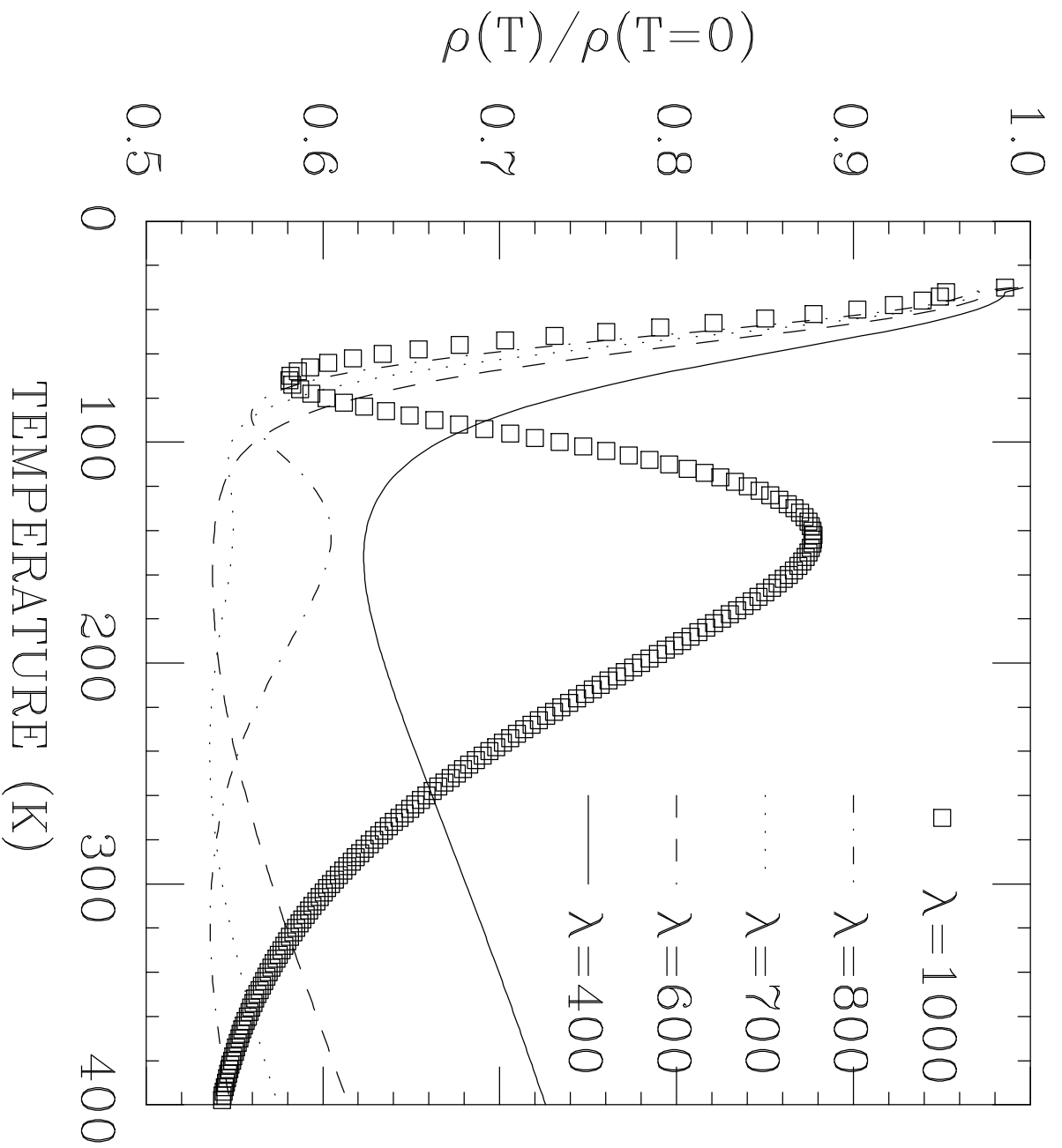
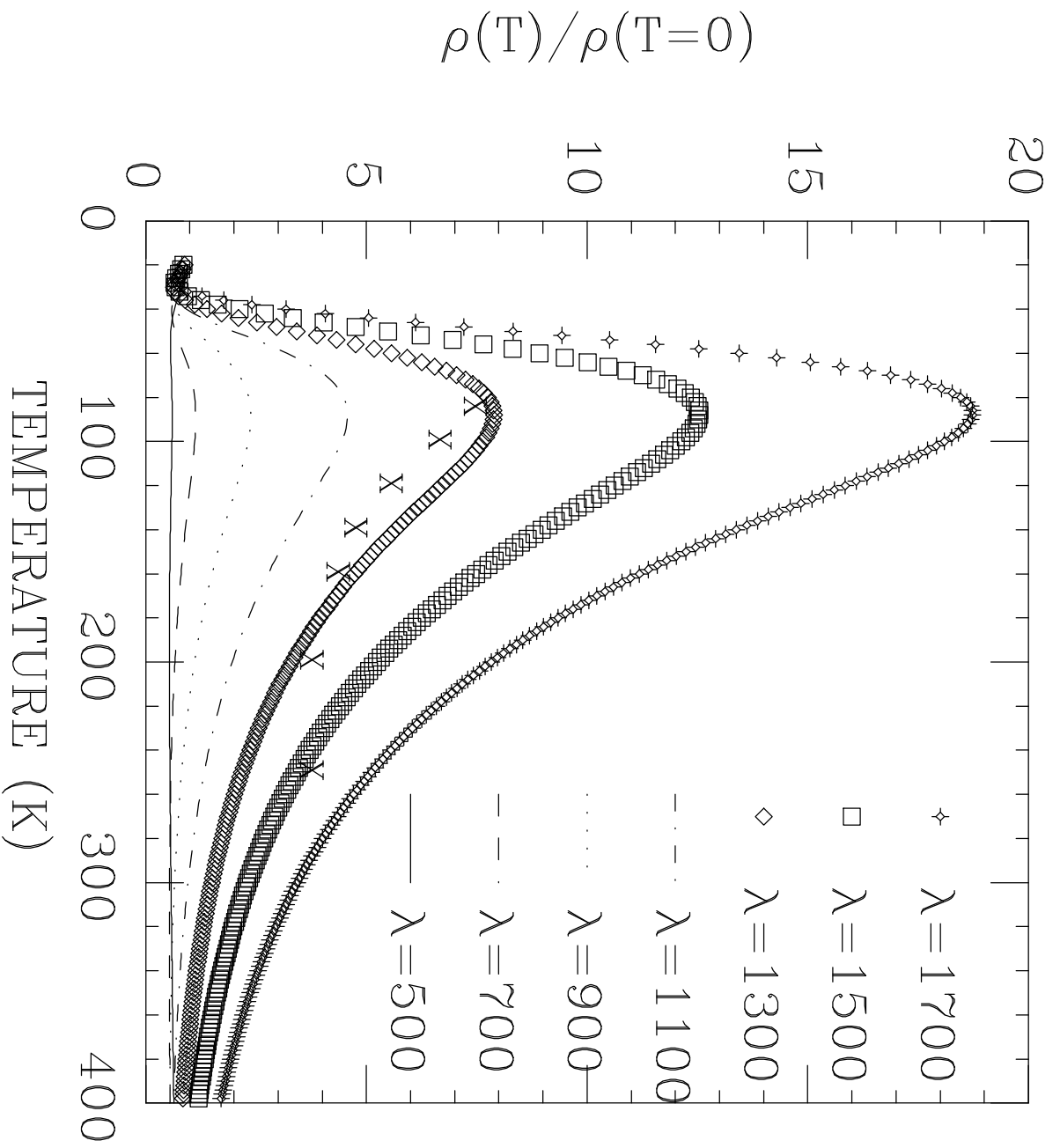
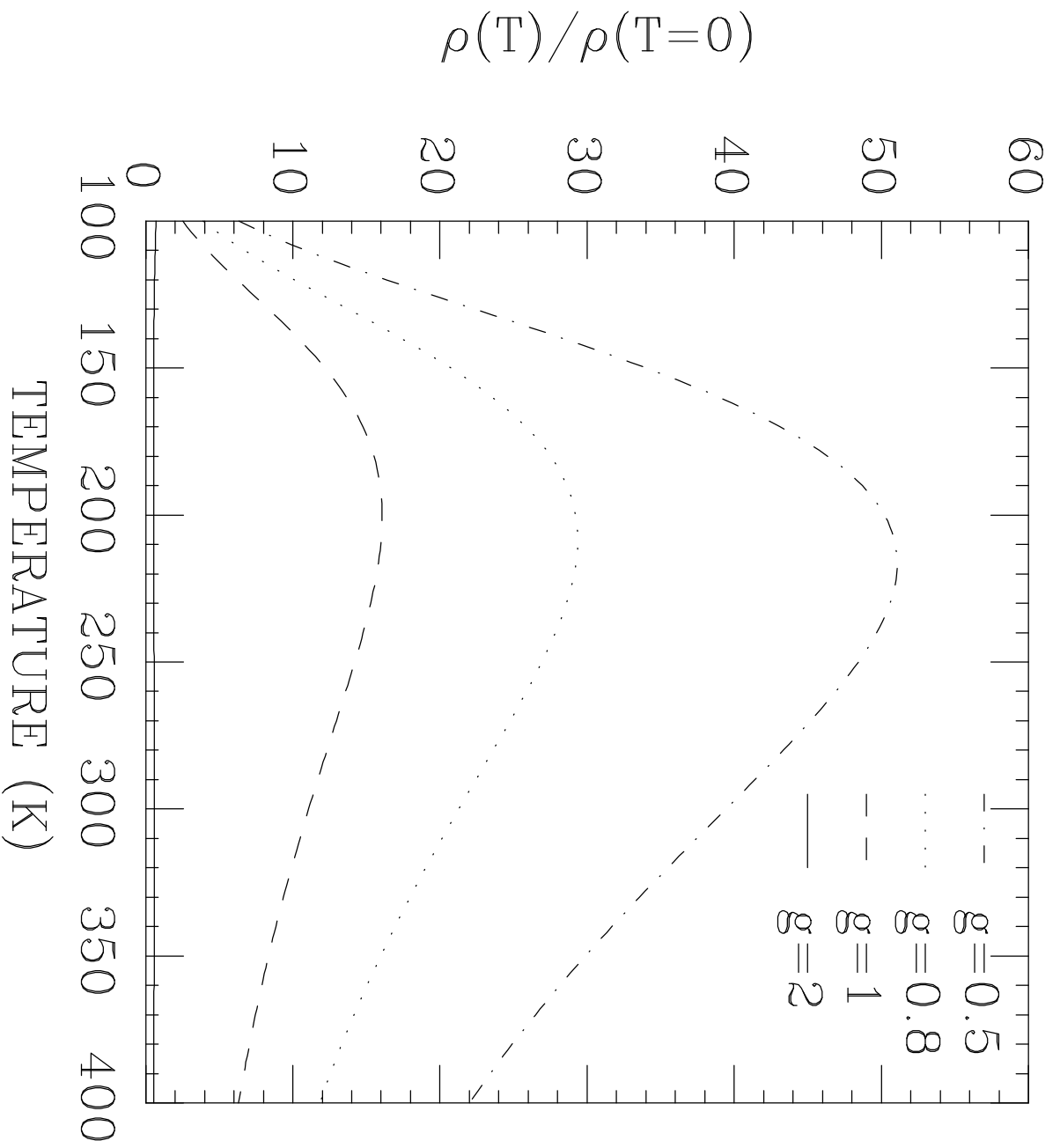
\end{document}